\documentclass[runningheads]{llncs}
\usepackage[T1]{fontenc}
\usepackage{graphicx}
\usepackage{placeins}
\usepackage{multirow}
\usepackage{subcaption}
\usepackage{amsmath}
\usepackage{amssymb}
\usepackage{algorithm}
\usepackage{algpseudocode}
\usepackage{float}
\usepackage{tikz}
\usepackage{float}
\usepackage[hidelinks]{hyperref}
\usepackage{color}

\begin{document}
\title{Transforming Single Photon Camera Images to Color High Dynamic Range Images}
%
\titlerunning{TSPCHDR}
%
\author{Sumit Sharma\inst{1}\orcidID{0009-0004-7454-3793} \and
Girish Rongali\inst{2}\orcidID{0009-0005-2239-6821} \and
Kaushik Mitra\inst{3}\orcidID{0000-0001-6747-9050}}
%
%
\institute{Indian Institute of Technology, Madras,India}
%
\maketitle              
\begin{abstract}
Traditional CMOS sensors suffer from restricted dynamic range and sub-optimal performance under extreme lighting conditions. They are affected by electronic noise in low light conditions and pixel saturation while capturing high illumination. Recent High Dynamic Range (HDR) imaging methods, often designed for CMOS sensors, attempt to address these issues by fusing multiple exposures. However, they frequently introduce artifacts like ghosting and light flickering in dynamic scenarios and non-uniform signal-to-noise ratios in extreme dynamic range conditions. Recently, Single Photon Avalanche Diode (SPAD) sensors, also known as Single Photon Camera (SPC) sensors, have surpassed CMOS sensors due to their capability to capture individual photons with high timing precision. Unlike traditional digital cameras that first convert light energy into analog electrical currents and then digitize them, SPAD sensors perform direct photon detection, making them less susceptible to extreme illumination conditions. Their distinctive non-linear response curve aids in capturing photons across both low-light and high-illumination environments, making them particularly effective for high-dynamic range imaging. Despite their advantages, images from SPAD sensors are often noisy and visually unappealing. To address these challenges, our paper evaluates state-of-the-art architectures for converting monochromatic SPAD images into color HDR images at various resolutions. We present a two-stage approach: first, transforming monochromatic SPAD images into color images via Image2Image translation networks, and second, enhancing these color images to achieve HDR via Single Image HDR methods. Our evaluation involves both qualitative and quantitative assessments of these architectures, focusing on their effectiveness in each stage of the conversion process.

\end{abstract}
\section{Introduction}
Traditional imaging systems often struggle with scenes featuring high levels of illumination and intricate details, such as bright light sources and highlights. Conventional image sensors have limited dynamic range because of limited full-well capacity. This leads to saturation at high brightness levels. To overcome the constraints of traditional sensors, several computational and hardware approaches have been developed \cite{reinhard2020high} for capturing High Dynamic Range (HDR) images. One common method is exposure bracketing, which combines multiple images taken at different exposure levels into a single HDR image \cite{debevec2023recovering},\cite{mann1994beingundigital}. However, this method can introduce ghosting artifacts \cite{tursun2015state} and light flicker in dynamic scenes \cite{velichko2017140}, \cite{iida20180}, \cite{deegan2018led}. 
To mitigate these issues, commercial HDR sensors typically use a limited number of exposures—usually between two and four—either sequentially or through dual-pixel architectures \cite{iida20180},\cite{asatsuma2019sub}. Despite these improvements, HDR cameras using CMOS sensors still face challenges such as motion artifacts and a reduced signal-to-noise ratio (SNR) in scenes with extreme dynamic ranges. The process of recovering an HDR image from a few exposures can lead to uneven SNR artifacts, making it difficult to preserve fine details and effective denoising of the image \cite{yang1999comparative}, \cite{mase2005wide}, \cite{velichko2017140}.

In contrast, Single Photon avalanche Diode (SPAD) sensors offer a promising solution by measuring brightness through precise timing of individual photons based on Poisson statistics. This approach allows SPAD sensors to surpass the saturation limits of traditional pixels and capture timing information with high accuracy \cite{ingle2021passive}. Recent advancements in Single-Photon Cameras (SPC) have demonstrated their superior dynamic range compared to conventional CMOS sensors \cite{liu2022single}. SPAD sensors excel in both low and high light conditions within a single exposure, enabling them to produce detailed images across a wide range of brightness levels \cite{ingle2019high}, \cite{ingle2021passive}. 
Despite these advantages, SPC images are not visually appealing, are noisy, and have limited resolutions. We propose two-stage as well as single stage approaches for converting monochromatic SPC images captured by single-photon cameras into high-resolution color HDR images. In the two-stage approach, we first utilize existing Image-to-Image translation networks to convert monochromatic SPC images to color LDR images and then convert them to color HDR images based on single image HDR reconstruction techniques. We evaluate various two-stage architectures through comprehensive quantitative and qualitative analyses.
We also evaluate a direct image2image translation method for converting monochromatic SPC images to color HDR images. Our main contributions are: 
\begin{itemize}
\item Transforming SPC images to color HDR images.
\item Detailed analysis of various two-stage architectures for achieving the above goal.
\item Single-stage Image2Image translation from monochromatic SPC images to color HDR images at high resolution produces comparable results.
\end{itemize}

\section{Related Work} 
\subsection{Image-to-Image Translation}
Given that Single Photon Camera (SPC) images lack color information, employing image-to-image translation methods presents a viable avenue for colorizing SPC images. Among these methods, Generative Adversarial Networks (GANs) stand out as a prominent framework for estimating generative models through an adversarial process. In this process, two models are concurrently trained: a Generative model 'G' that captures the data distribution, and a Discriminative model 'D' that estimates the probability that a sample originates from the training data rather than 'G' \cite{goodfellow2020generative}.
Building upon this concept \cite{goodfellow2020generative}, researchers have delved into conditional adversarial networks as a versatile solution to various image-to-image translation{I2I} challenges. These networks not only learn the mapping from input images to output images but also acquire a loss function to facilitate this mapping training \cite{isola2017image}. Defining and optimizing perceptual loss functions based on high-level features extracted from pre-trained networks makes it feasible to generate high-quality images \cite{johnson2016perceptual}. Consequently, the fusion of training feed-forward convolutional neural networks with per-pixel loss and perceptual loss functions has emerged as a favored approach for numerous image-to-image translation tasks \cite{dong2017semantic},\cite{kaneko2017generative},\cite{karacan2016learning}. Furthermore, various unsupervised methods have been proposed for tackling the image-to-image translation task \cite{bousmalis2017unsupervised},\cite{liu2017unsupervised},\cite{shrivastava2017learning,taigman2016unsupervised}.In contrast, one approach \cite{chen2017photographic} focuses on producing images with a photographic appearance that aligns with the input layout. This approach operates as a rendering engine, translating a two-dimensional semantic specification of the scene into a photographic image without relying on adversarial training, which may suffer from training instability and optimization issues. To circumvent these challenges, this approach \cite{chen2017photographic} adopts a direct regression objective based on computing distances between image features extracted by deep neural networks \cite{dosovitskiy2016generating}, \cite{gatys2016image}, \cite{johnson2016perceptual}. Consequently, although this approach \cite{chen2017photographic} yields the first model capable of synthesizing 2048 x 1024 images, it may lack finer details and textures.
Alternatively, another approach \cite{wang2018high} achieves visually appealing results at a resolution of 2048x1024 by incorporating a novel adversarial loss function, along with multi-scale generator and discriminator architectures. 
Unpaired image-to-image translation often relies on maximizing mutual information between source and translated images across different domains to preserve source content and prevent unnecessary modifications. Self-supervised contrastive learning has proven effective in this context, where it constrains features to ensure fidelity to the source. However, prior methods often use features from random locations to impose constraints, potentially lacking sufficient information from the source domain.
\subsection{Single Image HDR}
As a result of the non-linear response curve inherent in the SPAD pixel, characterized by an asymptotic saturation limit, and its efficacy in operating under exceptionally high flux levels, Single Photon Counting (SPC) devices exhibit the capability to capture High Dynamic Range (HDR) images even in challenging lighting conditions, such as those encountered in extremely low-light and high-light scenarios. Consequently, incorporating SPAD technology within single-image HDR methodologies is a promising avenue for achieving vibrant and color-rich HDR imagery \cite{ingle2019high},\cite{liu2022single}.
Reconstruction of high dynamic range (HDR) images through the fusion of multiple exposures stands as a conventional technique, harnessing the distinctive information encapsulated within each 
exposure\cite{debevec2023recovering},\cite{mertens2009exposure}.Specifically, low exposures encapsulate regions abundant in intensity, whereas high exposures capture details within low-intensity areas, a practice commonly referred to as Exposure Bracketing. Nonetheless, challenges arise within this methodology when tackled with motion scenarios, often resulting in artifacts such as ghosting. In such circumstances, Single-image HDR techniques present a promising alternative that is particularly adept at addressing issues like misalignment.
Eilertsen et al. \cite{eilertsen2017hdr} introduced a novel approach for reconstructing HDR images from low dynamic range (LDR) input images by inferring missing information within luminance-rich segments, such as highlights, which are lost due to limitations of the camera sensor. This was achieved through a fully convolutional neural network (CNN) structured in the form of a hybrid dynamic range autoencoder. Similarly, Yang et al. \cite{yang2018image} conceptualized the image correction task as an HDR transformation process, initially reconstructing absent details within the HDR domain, followed by tone-mapping to generate the output LDR image with reinstated details.
In contrast, Moriwaki et al. \cite{moriwaki2018hybrid} devised a method aimed at minimizing a hybrid loss, comprising perceptual and adversarial losses in addition to HDR reconstruction loss. Liu et al. \cite{liu2020single} modeled the HDR-to-LDR image formation pipeline, describing dynamic range clipping, non-linear mapping from a camera response function, and quantization. They further employed three specialized CNNs to reverse these steps, imposing effective physical constraints to facilitate the training of individual sub-networks. On the other hand, Santo et al. \cite{santos2020single} reconstructed HDR images by recovering saturated pixels from an input LDR image through a feature masking mechanism, thereby reducing the influence of features from saturated areas. To synthesize visually pleasing textures, they employed a VGG-based perceptual loss function.
The light-weight Deep Neural Network \cite{guo2022lhdr} has also been specifically designed to address noise and compression in Deep Neural Network-based Single Image High Dynamic Range (SIHDR) processing of legacy Standard Dynamic Range Images.
\section{Proposed work}
\subsection{Single Photon Camera Soft Saturation}
Conventional sensor pixels typically exhibit a linear input-output response, leading to hard saturation when subjected to a sudden increase in incident flux. In contrast, Single Photon Camera (SPC) pixels demonstrate a non-linear response characterized by an asymptotic saturation threshold. Upon detecting each photon, the Single Photon Avalanche Diode (SPAD) enters a predefined dead-time interval, rendering it temporarily insensitive to subsequent photons. This inherent property introduces non-linearity into the response curve of the SPAD, causing it to miss a fraction of incident photons as flux escalates in its passive mode.
The theoretical saturation limit of an SPC pixel in passive mode significantly surpasses that of conventional sensors, enabling it to resist saturation even under extreme brightness conditions. Instead of hard saturation, the SPC pixel reaches a soft saturation point where it remains responsive, albeit with a diminished signal-to-noise ratio. This soft saturation threshold notably exceeds the saturation limits observed in conventional sensors, making SPC pixels particularly well-suited for applications demanding extremely High Dynamic Range Imaging capabilities. 
\subsection{Single Photon Camera Simulator}
\begin{figure}[hbt!]
  \centering
  \includegraphics[width=8cm,height=15cm]{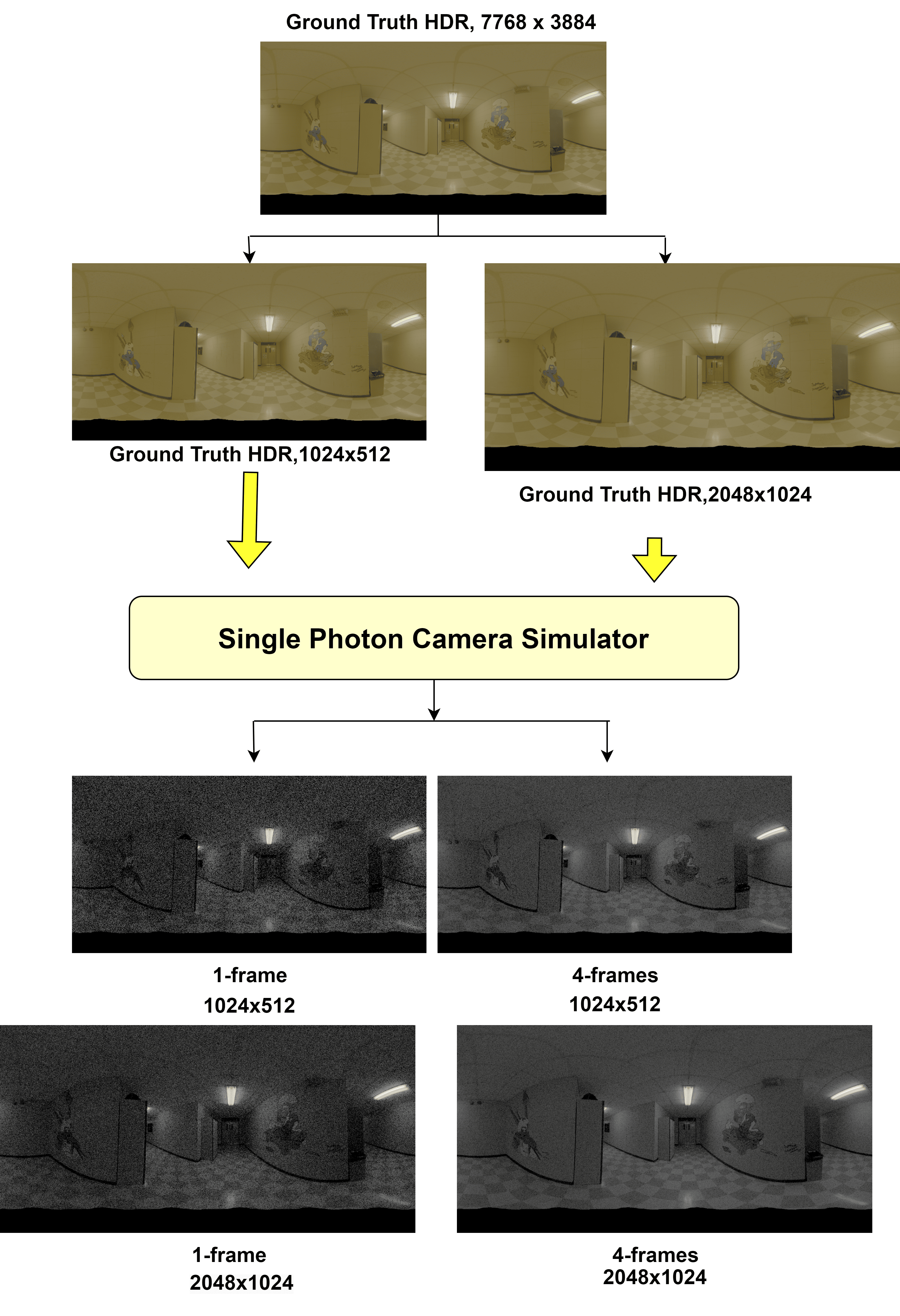}
  \caption{SPC simulation pipeline: Different averaged SPC frames at different resolutions are generated from the SPC simulator based on luminance-based exposure time selection.}
  \label{fig:simulator}
\end{figure}

Due to the continuous advancement of research in SPC and their functionalities, their availability in the commercial market remains limited. Hence, we prepare a Single Photon Camera simulator leveraging the insights from recent studies \cite{goyal2021photon},\cite{liu2022single}. This simulator is designed to replicate SPC image generation by employing ground-truth HDR images\cite{bolduc2023beyond} as input data. Essentially, it replicates real-life scene dynamics as if captured by a single-photon camera. The SPC Simulator has been developed under the assumption of the SPC operating in passive mode. With each photon detection event, the SPAD undergoes a dead time interval, during which it remains inactive and does not capture any photons. It is assumed that the arrival of each photon incident per pixel follows Poisson statistics. Meanwhile, the number of detected photons '$N^{T}_{SPC}$' within a fixed exposure time '$T$' follows a renewal process, as detailed in Grimmett and Stirzaker's work on probability \cite{grimmett2020probability}. Consequently, the mean and variance are determined by the methodology outlined in Ingle and Proakis's publication \cite{ingle2019high} as: 
\begin{equation}
\label{1}
E[{N^{SPC}_{T}}] =  q_{SPAD}\phi T /[1 + q_{SPAD} \phi \tau_{d}]
\end{equation}
\begin{equation}
\label{2}
Var[{N^{SPC}_{T}}] = q_{SPAD}\phi T /[1 + q_{SPAD} \phi \tau_{d}]^3,
\end{equation}
where $\phi$ is the photon flux, that is, number of photons per sec, $q_{SPAD}$ is the quantum efficiency of the SPC camera. Finally, we also add an averaging step in order to generate different numbers of averaged frames\cite{goyal2021photon} in the '.hdr' format.
The simulation begins by loading the High Dynamic Range (HDR) image using the OpenCV library. Once the HDR image is loaded, it is converted to gray-scale to facilitate the estimation of the luminance present within the image. This luminance is subsequently utilized to determine the appropriate exposure time, denoted as {$\tau_d$}, for HDR images to ensure optimal visibility. The gray-scale-converted image is treated as a \textbf{photon flux} '$\phi$', which is then used to calculate the \textbf{mean photon count} and \textbf{variance photon count} in accordance with the equations \ref{1} and \ref{2} outlined in \cite{liu2022single}.
\newline
During simulation, adjustments are made to key parameters such as exposure time ($T$) and SPC pixel sensitivity ($q_{SPAD}$) while maintaining a fixed dead time. The objective is to ensure visibility of all regions simulated from the SPC simulator, which would facilitate the attainment of an optimal reconstructed HDR image \cite{liu2022single}. In order to get the optimal results, we select the exposure time based on the luminance of the ground truth HDR images. Different resolution of simulator outputs can be seen from the \ref{fig:simulator}. We save the monochromatic SPC images obtained from the SPC simulator as 8-bit monochromatic images. This process was essential to prepare the input monochromatic SPC images for training and testing. 


\begin{figure*}[t!]
    \centering
    \includegraphics[width=\textwidth]{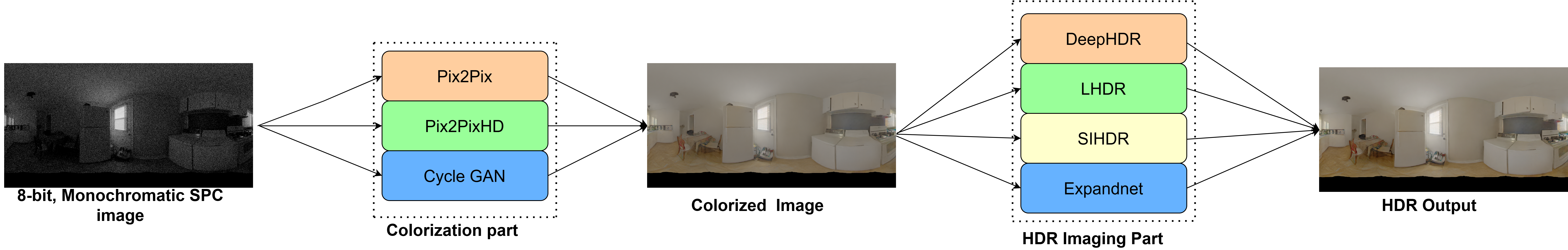}
    \caption{This study examines various methods for converting a monochromatic SPC image to a Color HDR image using state-of-the-art techniques. We explore Pix2Pix, Pix2PixHD, and CycleGAN for the translation from monochromatic to color. Subsequently, we investigate DeepHDR, LHDR, SIHDR, and ExpandNet for converting the colorized image to its HDR version.}
    \label{fig:Bin2hdr}
\end{figure*}
\begin{figure*}[t!]
    \centering
    \includegraphics[width=\textwidth]{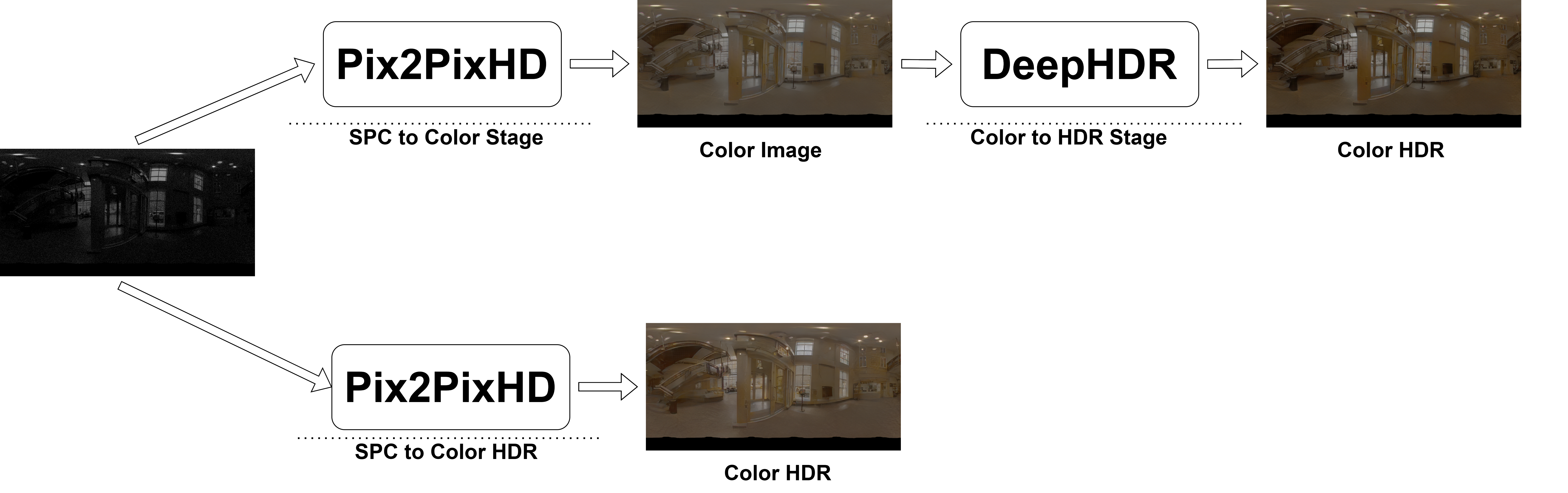}
    \caption{Best possible combination among the different state-of-the-art architectures. For image2image translation, we found Pix2PixHD performs best at low and high resolutions, and DeepHDR performs best when creating HDR versions among different state-of-the-art architectures. We also modified Pix2PixHD, for direct translation i.e., moving from Monochromatic HDR to Color HDR and getting Comparable results}
    \label{fig:Best possible combination}
\end{figure*}
\subsection{Two-stage and Single-stage approaches}
\label{sec:Directvsindirect}
SPC have limitations in resolution and color capabilities. However, Canon has developed a small SPAD sensor (13.2 mm x 9.9 mm) that can capture 3.2 megapixels of color images \cite{canon}. While there are potential hardware solutions, current prototypes are not ready to compete directly with CMOS image sensors used in smartphones. These sensors have strict requirements for pixel size, power use, and processing efficiency, which current SPAD technology does not yet meet \cite{ma2023seeing}.

This paper explores different computational imaging techniques to convert SPC monochromatic images into color high dynamic range (HDR) images. We explore two approaches for converting monochromatic SPC images to color HDR images: i) two-stage models and ii) single-stage models. In two-stage models, we explore different combinations of image-to-image SPC colorization methods and single-image HDR reconstruction methods. This is shown in Figure \ref{fig:Bin2hdr}. We explore Pix2PixHD\cite{wang2018high},Pix2Pix\cite{isola2017image} and CycleGAN\cite{zhu2017unpaired} for colorization task and  Expandnet\cite{marnerides2018expandnet}, DeepHDR\cite{santos2020single}, SIHDR\cite{liu2020single} and LHDR\cite{guo2022lhdr}, for single image HDR reconstruction. From Figure \ref{fig:Bin2hdr}, it is clear that a total of 12 combinations is possible to move from a SPC monochromatic image to a colorful HDR version.

To enhance the clarity of details in monochromatic SPC images such as text, we extended our study by applying averaging techniques. Specifically, we experimented with averaging different frames ranging from 1 to a maximum of 4. 
We create a paired dataset consisting of SPC simulated images, which are different frames averaged monochromatic SPC images  and corresponding logarithmic tonemapped color images of different resolutions, i.e., 1024x512 and 2048x1024 pixels.  We train the different image2image translation networks on these paired datasets for monochromatic SPC to color conversion. For the subsequent stage of color to HDR conversion, we used pre-trained models of various Single Image HDR methods.

Finally, we also tried single-stage translation to move from monochromatic image to color HDR image. For this, we trained Pix2PixHD\cite{wang2018high} between monochromatic SPC image and 32-bits-per-channel ground truth HDR images. The synthesized images are normalized images in the range 0-1, which are inverse tone-mapped images with gamma correction. We then post-process those images to 32-bits floating point images. These images were directly compared with 32-bits-per-channel ground truth HDR images in terms of HDR VDP scores.

\section{Experimentation}
All the experiments are done on the Laval indoor dataset \cite{gardner2017learning} for our analysis,  which contains a total of 2362 high-resolution HDR images of resolution 7768x3884. The images are downsampled to a resolution of 1024x512 and 2048x1024 to reduce the computational complexity. We selected these two different resolutions to assess performance improvements relative to image resolution. 
Monochromatic SPC frames, simulated from these downsampled images, serve as the input to the model. Since the SPC frames are very noisy, we also averaged a few frames (for example 4 frames) and gave them as input to the model, which we referred to as 4-frame averaged SPC images. 
As image translation methods cannot predict high bit-depth images directly, the monochromatic SPC images are tone-mapped to produce 8-bit monochromatic images, which are the input to the two-stage models. For training the colorization models in the two-stage models, we tonemapped the original color HDR images to produce color LDR images, which are then used as ground-truth for training the colorization networks. 
After extensive experimentation with Reinhard \cite{reinhard2020high} and Adaptive logarithmic tone mapping, we selected logarithmic tone mapping due to its superior performance.
For the second stage of the two-stage approach, we use different pre-trained state-of-the-art single-image HDR reconstruction models to convert the colorized SPC images to color HDR images.
All the experimentations are done using two NVIDIA GeForce RTX 3090 GPUs.


\section{Quantitative and Qualitative Analysis}
We perform quantitative analysis for both the stages of the two-stage apporaches: i) monochromatic SPC to colorized SPC and ii) colorized SPC to color HDR. For the task of SPC colorization, we used \textit{psnr},\textit{ssim}, and \textit{lpips} as metrics. For the task of converting color SPC to color HDR images, we use \textit{HDR-VDP3} \cite{mantiuk2023hdr} metric.

Table \ref{tab:quantitative} shows the comparison of various image-to-image translation models such as Pix2PixHD\cite{wang2018high}, Pix2Pix\cite{isola2017image}, and CycleGAN\cite{zhu2017unpaired} for the task of SPC colorization. The table illustrates the results with single frame and 4-frame (averaged) for 1024x512 and 2048x1024 resolutions. From the table \ref{tab:quantitative}, it is clear that Pix2PixHD \cite{wang2018high} achieves the highest performance for the colorization task among state-of-the-art methods, as measured by LPIPS, PSNR, and SSIM scores. Averaging 4 frames yields better results than using a single SPC monochromatic frame. 

For the task of generating HDR color images, we utilized single-image HDR reconstruction methods, including DeepHDR\cite{santos2020single}, SIHDR\cite{liu2020single}, ExpandNet\cite{marnerides2018expandnet}, and LHDR\cite{guo2022lhdr}. From Table \ref{tab:HDRVDP}, it is clear that the combination of Pix2PixHD\cite{wang2018high} and DeepHDR\cite{santos2020single} is quantitatively giving the best possible HDR reconstruction with the highest HDR-VDP3 scores among all methods. Also, the performance of these methods is better at higher resolutions of 2048x1024.

\begin{figure*}[htbp] 
    \centering
    \includegraphics[width=\textwidth]{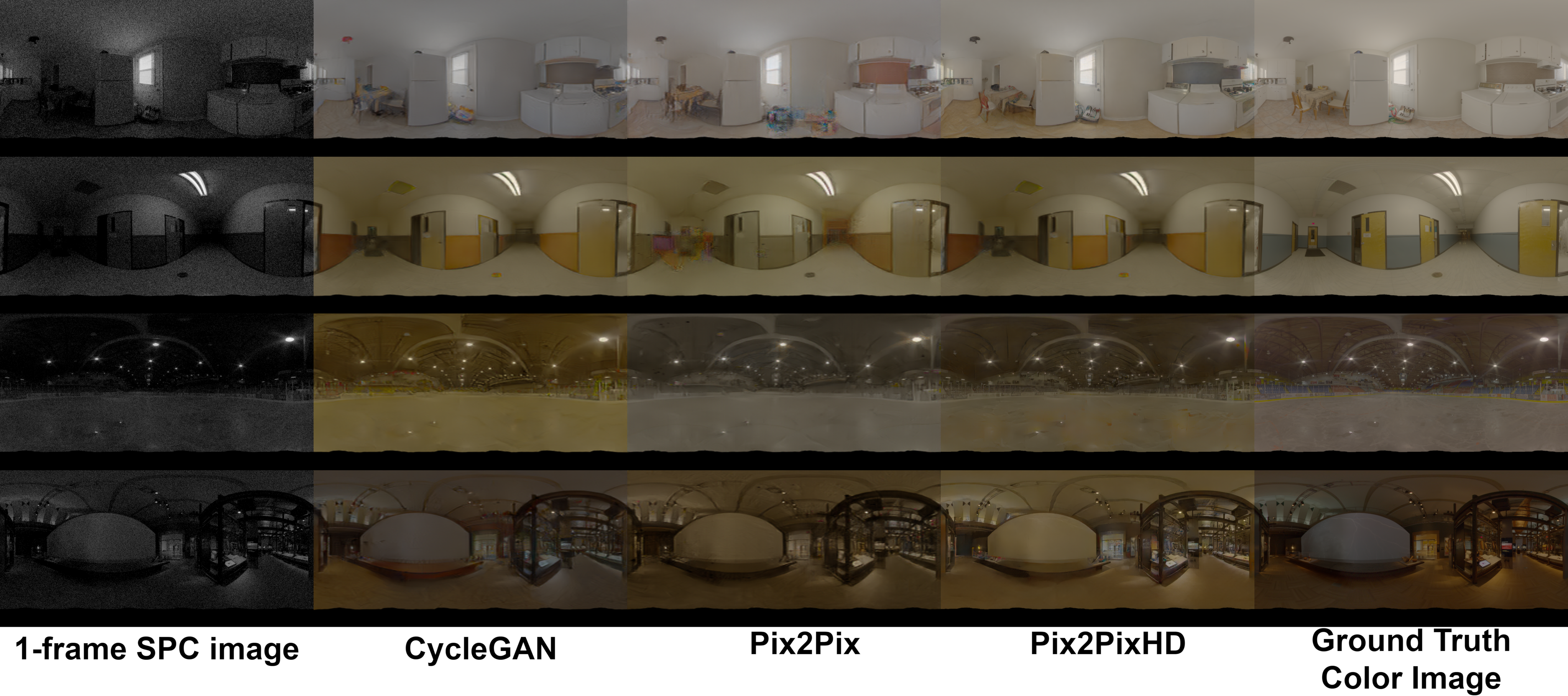} 
    \caption{This figure presents a qualitative analysis of 1-frame SPC to color translation at a resolution of 2048x1024. It displays various colorized outputs from image-to-image translation methods, with the rightmost column showing the logarithmic tone-mapped ground truth color images. Notably, Pix2PixHD yields the most favorable results. Please zoom in for further detail.}
    \label{fig:1frame_color}
\end{figure*}

\begin{figure*}[htbp] 
    \centering
    \includegraphics[width=\textwidth]{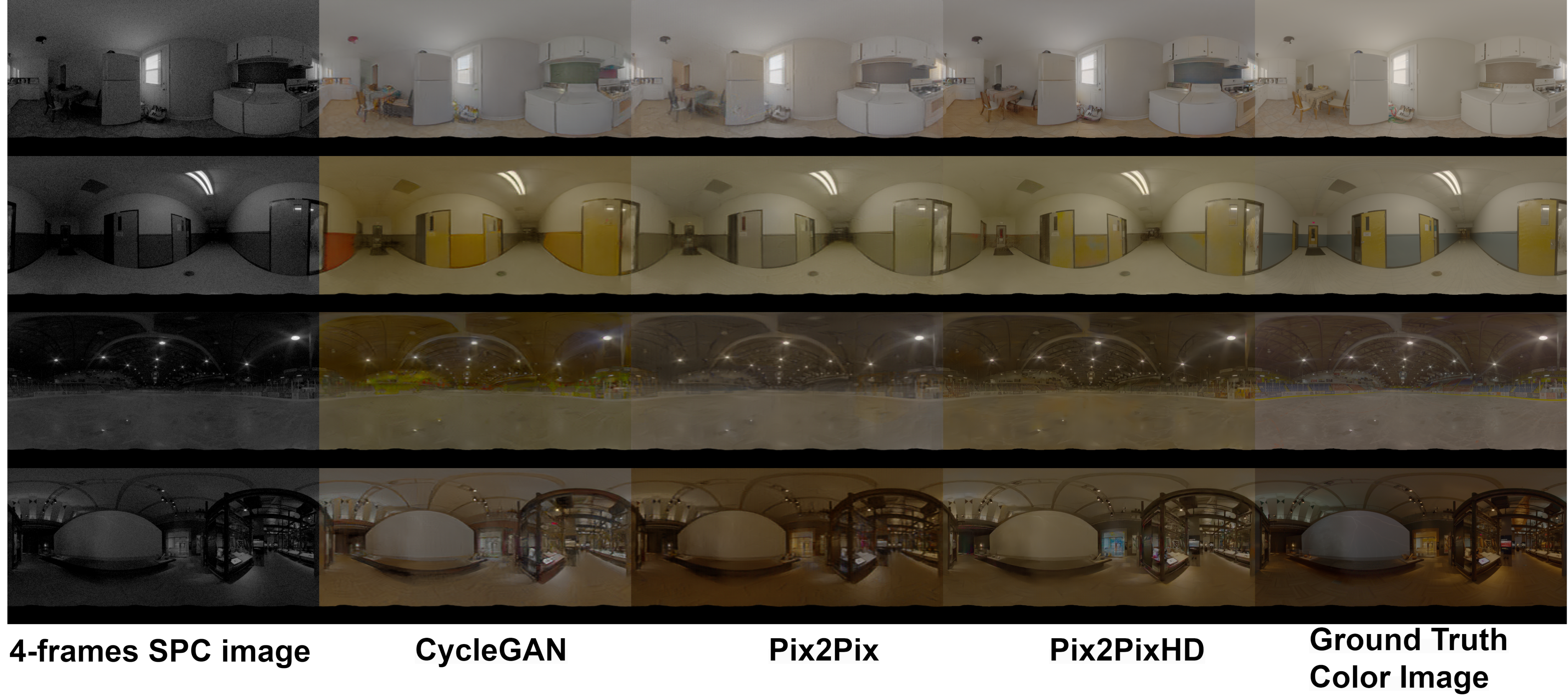} 
    \caption{This figure presents qualitative analysis of four-frame averaged SPC for color translation at a resolution of 2048x1024. It displays various image-to-image translation outputs. The rightmost column features logarithmic tone-mapped ground truth color images. Notably, the best results are achieved using Pix2PixHD.Please zoom in for further details}
    \label{fig:4frames_color}
\end{figure*}

Qualitatively, by observing the reconstructed images from different image-to-image translation networks in Figure \ref{fig:4frames_color}, it is clear that Pix2PixHD\cite{wang2018high} produces better results than other state-of-the-art architectures. We can also observe from figure \ref{fig:1frame_color} that CycleGAN\cite{zhu2017unpaired} fails to preserve colors and also fails to reconstruct the light sources faithfully at higher resolutions.

\begin{figure*}[htbp] 
    \centering
    \includegraphics[width=\textwidth]{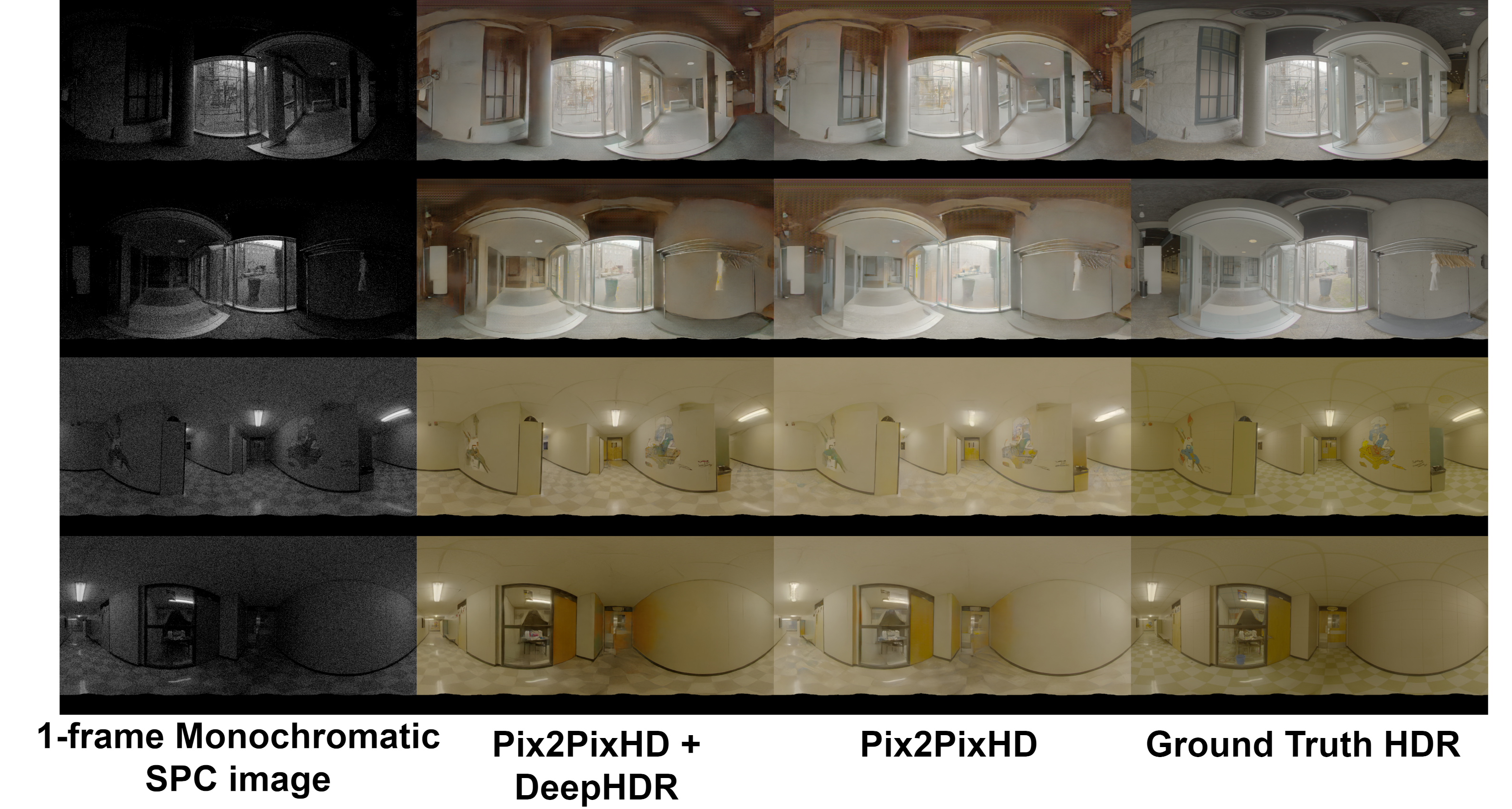} 
    \caption{Qualitative analysis of 1-frame SPC for color HDR translation at a resolution of 2048x1024 is presented. This figure displays outputs from both the two-stage (Pix2PixHD + DeepHDR) and single-stage (Pix2PixHD) methods. The rightmost column shows the logarithmic tone-mapped ground truth HDR images. Results demonstrate comparability between the approaches.Please zoom in for further details}
    \label{fig:color2hdr_1frame}
\end{figure*}

We also experimented with Pix2PixHD \cite{wang2018high} for the task of single-stage translation from Monochromatic SPC images to Color HDR images. Given that Pix2PixHD produced superior results among all models for the two-stage translation task, we modified its architecture as described in \ref{sec:Directvsindirect} for that task of single-stage translation.
As shown in Table \ref{tab:Direct_translation}, the results obtained using a single-stage translation task are slightly comparable but less than those obtained using two-stage translation. Figure \ref{fig:color2hdr_1frame} illustrates that the colors in the HDR image produced by the two-stage translation (Pix2PixHD with DeepHDR) are closely aligned with the ground truth, whereas the colors generated by the modified Pix2PixHD in the single-stage translation deviate slightly from the ground truth.Please refer to Figure \ref{fig:Best possible combination} for a comparison of direct and indirect approaches, illustrating the optimal combination of these methods. Single Image HDR techniques are primarily designed to convert Low Dynamic Range (LDR) images to High Dynamic Range (HDR) images, which typically involves expanding the bit depth of each color channel. However, for the task of translating monochromatic Single-Photon Camera (SPC) images to color HDR images, additional losses are required to effectively predict the color information. In this context, \textbf{adversarial loss} can be a promising approach \cite{wang2018high}.   


\begin{table*}[hbt!] 
    \caption{SPC monochromatic to colorization results. Pix2PiHD gives the best result.}
    \label{tab:quantitative}
    \centering
    \resizebox{\textwidth}{!}{
    \begin{tabular}{|c|c|c|c|c|c|c|c|}
      \hline
      Method & Resolution & lpips 1-frame$\downarrow$ & lpips 4-frames$\downarrow$ & PSNR 1-frame$\uparrow$ & PSNR 4-frames$\uparrow$ & SSIM 1-frame$\uparrow$ & SSIM 4-frames$\uparrow$ \\
      \hline
      \textbf{Pix2PixHD}\cite{wang2018high} & 1024x512 & \textbf{0.2417} & \textbf{0.2029} & \textbf{25.34} & \textbf{25.36} & \textbf{0.8456} & \textbf{0.8718} \\
      \hline
      Pix2Pix\cite{isola2017image} & 1024x512 & 0.4102 & 0.3901 & 24.87 & 25.09 & 0.8279 & 0.848 \\
      \hline
      CycleGAN\cite{zhu2017unpaired} & 1024x512 & 0.4068 & 0.3973 & 23.1 & 23.25 & 0.8192 & 0.8235 \\
      \hline
      \textbf{Pix2PixHD}\cite{wang2018high} & 2048x1024 & \textbf{0.2254} & \textbf{0.1995} & \textbf{26.53} & \textbf{26.5} & \textbf{0.887} & \textbf{0.9056} \\
      \hline
      Pix2Pix\cite{isola2017image} & 2048x1024 & 0.4418 & 0.443 & 25.25 & 25.23 & 0.8641 & 0.8701 \\
      \hline
      CycleGAN \cite{zhu2017unpaired} & 2048x1024 & 0.4498 & 0.4547 & 23.2 & 23.07 & 0.8384 & 0.8364 \\
      \hline
    \end{tabular}}
\end{table*}

\begin{table*}[hbt!]
\centering
\caption{SPC monochromatic to HDR results. The HDR VDP scores are averaged scores over 262 images. Pix2PixHD + DeepHDR gives the best result.}
\label{tab:HDRVDP}
\resizebox{\textwidth}{!}{%
\begin{tabular}{|c|c|c|c|c|c|c|c|}
\hline
\multirow{2}{*}{Method} & \multirow{2}{*}{Resolutions}& \multicolumn{2}{c|}{Q score (1-frame) $\uparrow$} & \multicolumn{2}{c|}{Q score (4-frames) $\uparrow$} \\
\cline{3-6}
& &1024x512&2048x1024&1024x512&2048x1024 \\
\hline
Pix2PixHD\cite{wang2018high} + Expandnet \cite{marnerides2018expandnet} &Low+High&3.1596&3.41&3.2667& 3.4533 \\
\hline
\textbf{Pix2PixHD\cite{wang2018high} + DeepHDR\cite{santos2020single}} &Low+High& \textbf{3.6388} & \textbf{3.9078} & \textbf{3.7393} & \textbf{4.0079} \\
\hline
Pix2PixHD\cite{wang2018high} + SIHDR \cite{liu2020single} &Low+High&2.7251&3.1357&2.824&3.2111\\
\hline
Pix2PixHD\cite{wang2018high} + LHDR \cite{guo2022lhdr} &Low+High&3.1131&3.386&3.2056&3.4339\\
\hline
Pix2Pix\cite{isola2017image} + Expandnet \cite{marnerides2018expandnet} &Low+High&3.0453&3.1783&3.1666&3.2977 \\
\hline
Pix2Pix\cite{isola2017image} + DeepHDR \cite{santos2020single} &Low+High&3.4545&3.5753& 3.5624&3.6699\\
\hline
Pix2Pix\cite{isola2017image} + SIHDR \cite{liu2020single} &Low+High&2.8896&3.0503&3.0131&3.1685 \\
\hline
Pix2Pix\cite{isola2017image} + LHDR \cite{guo2022lhdr} &Low+High&3.1354&3.2717&3.2344&3.3639 \\
\hline
CycleGAN\cite{zhu2017unpaired} + Expandnet \cite{marnerides2018expandnet} &Low+High&2.9679&3.1897&3.1041&3.2136 \\
\hline
CycleGAN\cite{zhu2017unpaired} + DeepHDR \cite{santos2020single} &Low+High&3.3189&3.5239&3.4861&3.581 \\
\hline
CycleGAN\cite{zhu2017unpaired} + SIHDR \cite{liu2020single} &Low+High&2.7932&3.0345&2.9699&3.1025 \\
\hline
CycleGAN\cite{zhu2017unpaired} + LHDR \cite{guo2022lhdr} &Low+High&3.0445&3.2358&3.118& 3.2432 \\
\hline
\end{tabular}%
}
\end{table*}
\begin{table}[h!]
\caption{Single stage translation from SPC monochromatic to color HDR}
\centering
\begin{tabular}{|c|c|c|c|}
\hline
Method&Resolution&1-frame,Q score&4-frames,Q score\\
\hline
Pix2PixHD&1024x512&3.4868&3.6154 \\
\hline
Pix2PixHD&2048x1024&3.8355&3.9734\\
\hline
\end{tabular}
\label{tab:Direct_translation}
\end{table}

\section{Conclusion and future works}
This paper presents a detailed analysis of transforming single-photon camera (SPC) captured images into colorful HDR images using two-stage architecture. SPC images inherently possess high dynamic range due to their soft-saturation characteristics \cite{liu2022single}, \cite{ingle2021passive}, \cite{ingle2019high}. Our simulator \cite{liu2022single} is capable of generating single-frame monochromatic HDR images by adjusting parameters such as SPC sensitivity, exposure time, and a \textbf{fixed dead time} of \textbf{150 ns}. The exposure time is determined based on the \textbf{luminance} present in the image. Our two-stage approach involves image-to-image translation methods for colorization and single-image HDR reconstruction methods for generating the final HDR version. We have also compared the two-stage approach with a single-stage approach and obtained comparable results. 

Given that single photon cameras can record high frame rates exceeding 96k frames per second, this research can be extended to develop high-frame-rate HDR videos, which will enhance our understanding of various scenes.
\section{Acknowledgement}
This work was supported in part by IITM Pravartak Technologies Foundation. 

\FloatBarrier

\end{document}